# Terahertz Kerr effect

Matthias C. Hoffmann, Nathaniel C. Brandt, Harold Y. Hwang , Ka-Lo Yeh, and Keith A. Nelson

*Massachusetts Institute of Technology, Cambridge, MA 02139*

We have observed optical birefringence in liquids induced by single-cycle THz pulses with field strengths exceeding 100 kV/cm. The induced change in polarization is proportional to the square of the THz electric field. The time-dependent THz Kerr signal is composed of a fast electronic response that follows the individual cycles of the electric field and a slow exponential response associated with molecular orientation.

The Kerr effect is a change $\Delta n$ in the optical refractive index which is quadratic in the externally applied electric field. In the DC limit, this is usually expressed as $\Delta n = K\lambda E^2$ with the Kerr constant $K$ and the vacuum wavelength $\lambda$. At optical frequencies an intensity-dependent modulation of the refractive index $\Delta n = n_2^I I(t)$ is observed, resulting in well-known nonlinear optical effects like self-focusing (Kerr-lensing), self-phase modulation, and birefringence that is usually measured through the depolarization of a separate, weak optical beam in what is conventionally designated the optical Kerr effect (OKE) measurement. The optical Kerr effect [1-4] can be observed in isotropic materials including liquids and gases that do not possess a second-order contribution $\chi^{(2)}$ to the polarization that leads to a linear dependence on the electric field (Pockels effect). Using ultrashort laser pulses, information about the dynamical responses of liquids on femtosecond time scales can be obtained by optically heterodyned pump-probe measurements of transient birefringence (OHD-OKE) [5-7]. In addition to orientational degrees of freedom, molecular vibrations excited through impulsive stimulated Raman scattering may be observed.

Here we present results in which the refractive index at optical frequencies is perturbed not by optical pulses but single-cycle terahertz (THz) pulses. Processes in the terahertz frequency range correspond to timescales on the order of molecular relaxation constants in liquids and glasses, but until recently the THz field strengths have been too small to induce observable higher-order nonlinear effects. Due to the phase-stability of THz generation processes, a femtosecond optical pulse that is used to produce a THz pulse can be used to time-resolve individual cycles of the THz field, and we anticipate that the optical pulse also can be used to time-resolved a Kerr response to individual cycles of the THz field, in contrast to all-optical Kerr experiments where only the Kerr response to the intensity envelope of the optical excitation pulse can be observed. Z-scan measurements of ZnTe during THz generation by an optical pulse [8] suggested significant THz-induced Kerr index change, but conventional OKE and mixed optical/THz effects, as well as cascaded $\chi^{(2)}$ effects in the noncentrosymmetric crystal, were also present. Here we present results in centrosymmetric media with only THz excitation to demonstrate unambiguously and generally the THz Kerr effect.

The experimental setup is shown in Figure 1. Single-cycle THz pulses with energies exceeding 1.5 µJ were generated by the tilted pulse front technique [9-11]. This method uses noncollinear velocity matching to enhance optical rectification in lithium niobate through tilting of the intensity front of a femtosecond laser pulse with a grating. The generated field was collimated and focused onto the sample where the THz intensity exceeded 50 MW/cm$^2$. An Infrasil cuvette that is relatively transparent in the far-infrared, with a path length of $L$ = 5 mm, was used for liquid samples. CS$_2$, benzene and CCl$_4$ have low absorption in the THz range, so a 5 mm sample length did not significantly diminish the THz pulse.

A weak 800 nm probe beam was passed collinearly through the sample at a polarization of 45 degrees with respect to the THz polarization. A combination of a quarter-wave plate and a Wollaston prism was used to analyze the change in refractive index. Two balanced photodiodes and a lock-in amplifier were used to record pump-probe data. In order to assess the field strength and temporal shape of the THz field at the sample position, electro-optic sampling [12] with a 0.1 mm ZnTe crystal was used.

Since in our experiment, the THz excitation pulse and the optical probe pulse travel through the sample together, it is important to achieve velocity matching. The time-domain signal is optimal if the phase velocity of the THz pulse causing the birefringence and the group velocity of the optical pulse probing it are identical. In this case the optical pulse envelope remains fixed at the same part of the THz wave cycle as the two pulses propagate through the thick sample. The matching of the optical (group) refractive index and the THz refractive index in our samples allows us to use long path lengths, thus enhancing measurement sensitivity to small refractive index changes.

Figure 2a shows results of THz-pump/optical-probe scans for $CS_2$, $CH_2I_2$, benzene, $CCl_4$ and chloroform, normalized to unity. The overall signal levels in the different samples reflect the differences in electronic and molecular polarizabilities. $CS_2$, $CH_2I_2$, and benzene have the strongest signals, which show slow decays due to the orientational contributions to their polarizabilities [5, 13] as will be discussed in more detail below. The observed signal from $CCl_4$ is substantially weaker and has no orientational component because of the molecular symmetry. $CHCl_3$ should show orientational signal but the level is apparently too low to detect above noise in our current measurements. THz Kerr effect responses were also observed from a solid sample, a 150-micron thick crystal of the relaxor ferroelectric $KTa_{1-x}Nb_xO_3$ (KTN) with x = 0.09 at room temperature, in the paraelectric phase. Velocity-matching was not possible in this material because of its high THz refractive index, and the signals, which showed a long decay time, are under further study. Surprisingly, nitrobenzene which has a value of $5.8\times10^{-20}$ m$^2$ V$^{-2}$ for $\chi^{(3)}$ at optical frequencies [14], roughly twice that of CS$_2$, yielded no measurable signal. This was also the case for tetrahydrofuran (THF), which is known to have a very low polarizability.

The magnitudes of the observed Kerr signals scale quadratically with the THz field amplitude (Figure 2b) as expected for a $\chi^{(3)}$ process. From the observed magnitude of the Kerr signal, i.e. from the difference signal $\Delta I/I$ in our balanced detection system, the phase retardation $\Delta\phi$ of the probe pulse of frequency $\omega$ accumulated while traversing the sample cell with length $L$ can be calculated and the change in refractive index $\Delta n$ caused by the THz field can be deduced from the expressions

$$\frac{\Delta I}{I} = \sin\Delta\phi = \sin\frac{\Delta n \omega L}{c} = \sin\frac{(n_x - n_y)\omega L}{c} \qquad (1)$$

At the maximum THz pump strength we obtain $\Delta I/I$ of $2.7\times10^{-3}$, corresponding to an index change of $\Delta n = 4.3\times10^{-6}$, in the case of CS$_2$. For a THz electric field of 150 kV/cm we obtain a Kerr constant $K$ of $2.4\times10^{-14}$ m/V$^2$ which is close to the DC value for CS$_2$ of $2.8\times10^{-14}$ m/V$^2$ [15]. In all-optical Kerr measurements, the refractive index change for a weak probe of linearly polarized light at frequency ω' is connected to the nonlinear refractive index $n_2$ by

$$\Delta n(\omega') = 2n_2 \langle E(\omega)E(\omega) \rangle \qquad (2)$$

and related to the nonlinear susceptibility by $n_2 = \frac{3}{2n_0}\chi^{(3)}_{xxxx}(\omega';\omega',\omega,-\omega)$ where $n_0$ is the unperturbed refractive index. The angled brackets indicate a time average over the electric field cycles. The nonlinear refractive index can also be given in terms of intensity $\Delta n = n_2^I I$ where the relationship between the nonlinear coefficients is given by $n_2^I = \dfrac{n_2}{\varepsilon_0 n_0 c}$ where $\varepsilon_0$ the vacuum permittivity. Table I summarizes our results for the observed peak refractive index change and the derived nonlinear constants for the five liquids.

Generally, our values for the nonlinear refractive index are on the same order of magnitude as values reported for all-optical measurements [14]. More quantitative comparison is difficult because in the available literature we found substantial variation in the reported values for $n_2$ and $\chi^{(3)}$ from OKE measurements.

The time-dependent polarization $P$ can be generally expressed using a response function $\mathbf{R}(t)$ [16]

$$P^{(3)}_{ijkl}(t,\tau) = E_j^{pr}(t-\tau)\int_0^\infty d\tau' R^{(3)}_{ijkl}(\tau') E_k^{pu*}(t-\tau') E_l^{pu}(t-\tau') \qquad (3)$$

where $\tau$ is the delay between the pump and probe pulses $E^{pu}$ and $E^{pr}$. The response function may be further separated into electronic and nuclear degrees of freedom $R^{(3)}_{ijkl} = R^{(3)el}_{ijkl} + R^{(3)nuc}_{ijkl}$. For electronically nonresonant frequencies, the electronic response function is essentially instantaneous in time and contains no information about molecular dynamics. The second term contains the inter- and intramolecular contributions to the signal. Since the frequency content of our pump THz pulses is smaller than 3 THz (100 cm$^{-1}$) we expect no contribution from intramolecular Raman-active vibrations which have higher frequencies in all our liquid samples. For anisotropic molecules the main contribution to the nuclear term is from molecular orientational diffusion, which can be approximated to have an exponential decay with time constant $\tau_0$. The induced difference $\Delta n$ between the indices of refraction parallel and perpendicular to the THz field can then be written as

$$\Delta n(\tau) = n_2^e E^2(\tau) + \frac{n_2^o}{\tau_o} \int_{-\infty}^{t} E_{THz}^2(t) \times e^{-(t-\tau)/\tau_o} dt' \qquad (3)$$

where $n_2^e$ contains the instantaneous the electronic contribution and $n_2^o$ the orientational contribution to $n_2$. A fit of Equation 3 to $\Delta n(t)$ for $CS_2$ yields ratio $n_2^o/n_2^e$ of about 0.4; subtraction of the electronic contribution from the measured response yields the red trace in Fig. 3a which gives the purely molecular orientational contribution with orientational relaxation time $\tau_0 = 1.7$ ps. In the case of diiodomethane ($CH_2I_2$), we observe a strong electronic contribution as well as a longer exponential contribution with a time constant of 13 ps (Fig 3b). This relatively large value can be explained by the larger moment of inertia of the diiodomethane molecule. The exponential decay times that we measured are in reasonable agreement with reported values from all-optical Kerr effect measurements [13].

In $CH_2I_2$ and $CCl_4$, it can be seen that on the short time scale the nonlinear refractive index follows the square of the THz field. This is in contrast to all-optical Kerr measurements, where the birefringence change follows the *envelope* of the square of the electric field. $CH_2I_2$ and $CCl_4$ are favorable samples for this observation because in the former the electronic Kerr response, which is the part that follows the square of the pump field, is well separated temporally from the much slower orientational response, while in the latter there is no orientational response. A more extensive theoretical model of dynamic responses to THz fields [17] may enable spectroscopic exploitation of the fact that induced responses with well defined polarities can be measured. In molecules like THF, nitrobenzene or water that have substantial molecular dipole moments, a significant orientational response should be induced through direct interaction with the THz field. However, orientational responses that are linear in the THz field will not be observed by the probe since this would be a $\chi^{(2)}$ measurement in an isotropic sample. A second THz field interaction with molecular dipoles might drive second-order orientational responses, but so far we have not observed measurable optical birefringence in polar liquids although THz-induced second-order lattice vibrational responses were observed optically in ferroelectric crystals [18]. Therefore a strong molecular dipole does not necessarily facilitate the present class of measurement, and may hinder it through absorption of the THz pulse near the front of the sample. Other probing methods may enable detection of nonlinear THz pumping of molecular dipolar responses.

We have observed THz-induced transient optical birefringence in liquids. The nonlinear refractive index $n_2$ is generally on the same order of magnitude as in all-optical measurements. In materials with a small or slow contribution of molecular orientation such as $CCl_4$ and $CH_2I_2$, we are able to observe that the electronic part of the system response follows roughly the square of the THz electric field. Therefore responses to each THz field polarity may be observed distinctly. This may enable novel spectroscopic measurements of responses with well defined orientations in addition to alignments. The THz Kerr effect may reveal polarizability dynamics associated with electronic, vibrational, and structural responses in ordered and disordered solids as well as liquids.

Table I: Refractive index change $\Delta n$, nonlinear refractive index $n_2$ in terms of intensity, nonlinear susceptibility $\chi^{(3)}(\omega_{THz},\omega_{THz},\omega_{opt})$ and rotational relaxation time constant $\tau_0$ for the THz Kerr effect in various liquids. The measured THz pulse parameters used for the calculation are: duration 1 ps, beam diameter at focus 2 mm, energy 1.5 μJ. These correspond to a THz peak field of 150 kV/cm and peak intensity of 50 MW/cm$^2$ The fourth column contains reference values from [14]

| Liquid | $\Delta n$ ×10$^{-6}$ | $K$ (10$^{-14}$ m/V) | $n_2^I$ (10$^{-16}$ cm$^2$/W) | $n_2^I$ [14] (10$^{-16}$ cm$^2$/W) | $\chi^{(3)}$ (10$^{-20}$ m$^2$/V$^2$) | $\tau_0$ (ps) |
|---|---|---|---|---|---|---|
| CS$_2$ | 4.3 | 2.4 | 440 | 332 | 2.08 | 1.7 |
| Benzene | 0.5 | 0.26 | 56 | 168 | 0.22 | 2.1 |
| CCl$_4$ | 0.23 | 0.12 | 27 | 15 | 0.10 | - |
| CHCl$_3$ | 0.086 | 0.045 | 10 | 30 | 0.04 | - |
| CH$_2$I$_2$ | 1.4 | 0.75 | 140 | 147 | 0.70 | 13 |

**Figure 1 (color online): (a)** Experimental setup. THz radiation was generated by the tilted pulse front method and focused onto the sample in a 5 mm quartz cell. Light pulses with 800 nm wavelength and 100 fs duration were used to probe the sample, and their induced depolarization was analyzed. **(b)** The probe polarization at the sample was 45 degrees with respect to the THz polarization.

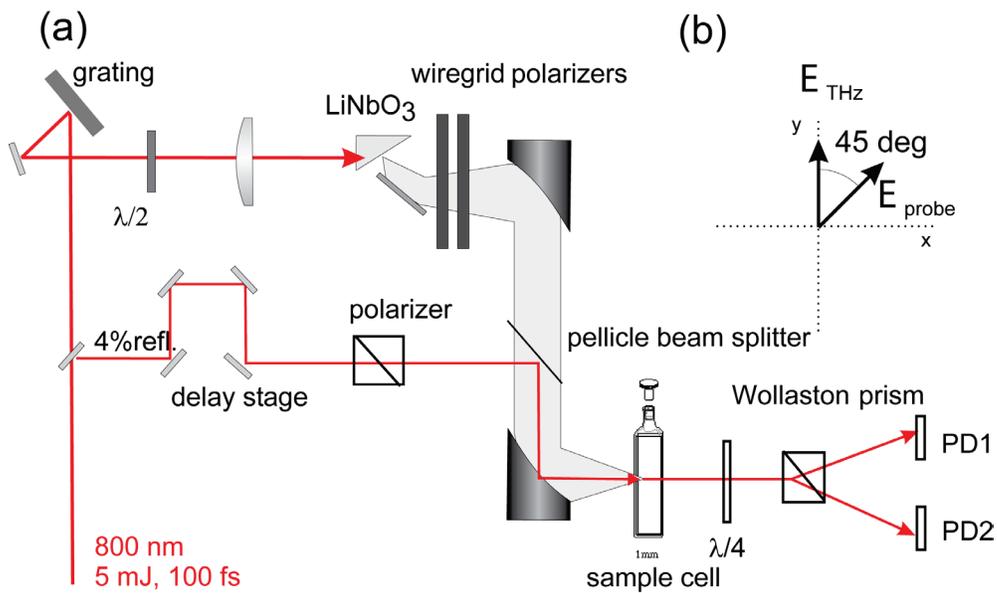

**Figure 2 (a)** THz-Kerr signals obtained from five different liquids. The dotted line indicates the square of the electric field measured by electro-optic sampling with ZnTe. **(b)** The magnitude of the Kerr signal (shown for $CS_2$) scales quadratically with the applied THz field.

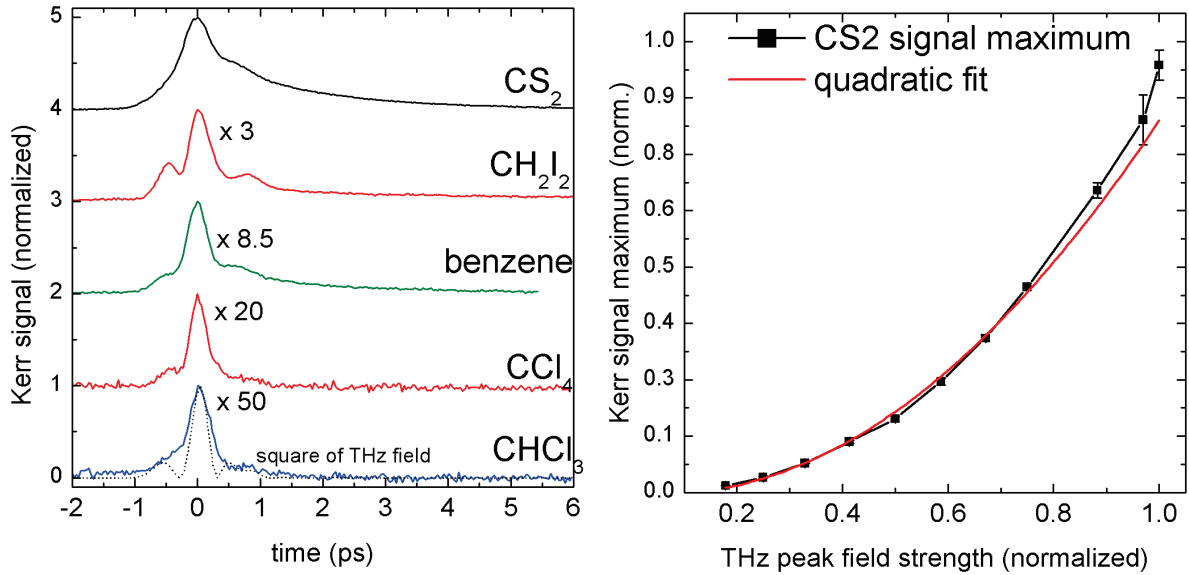

**Figure 3:** (color online) THz Kerr signal in (a) $CS_2$, (b) $CH_2I_2$ and (c) $CCl_4$ (solid lines). For comparison, the square of the THz electric field profile obtained from EO sampling is shown (dotted lines). The dashed (blue) line in (a) is based on a fit to equation 4. The insets show data on a log scale together with exponential fits yielding decay constants of 1.7 ps in $CS_2$ and 13 ps in $CH_2I_2$.

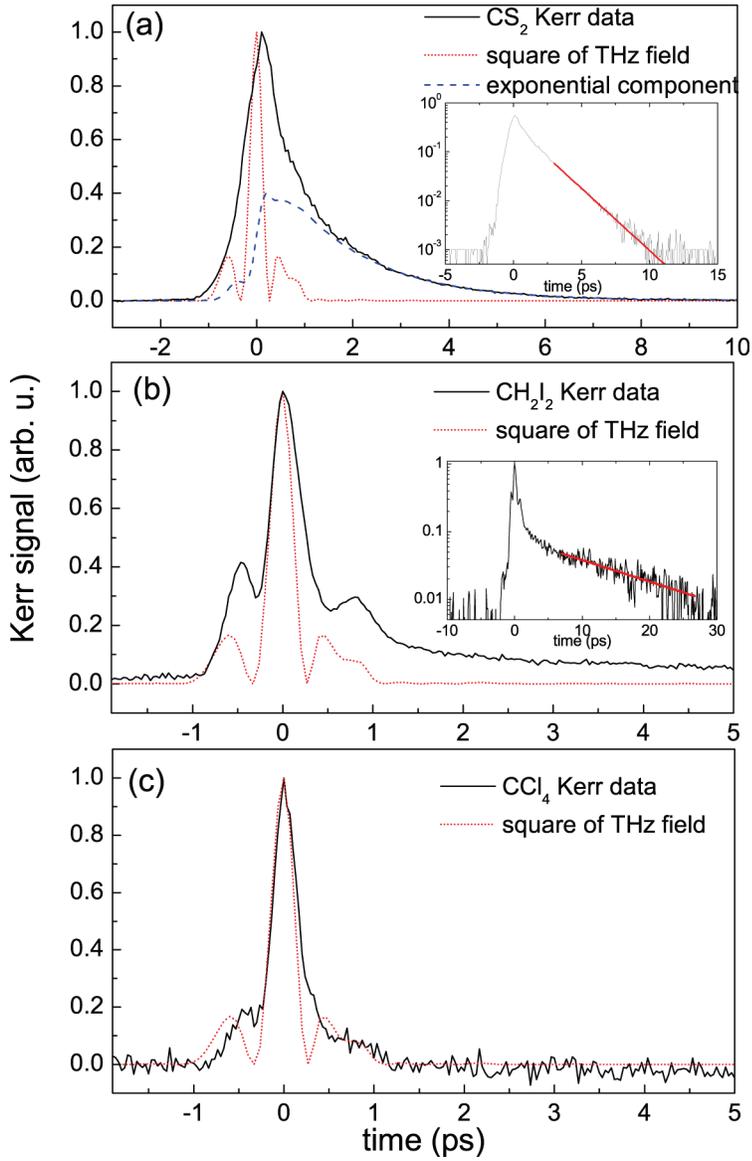